\documentclass[cits]{PoS}
\usepackage{url}
\usepackage{amsmath}
\usepackage{listings}
\title{Fast Multi-tone Phase Calibration Signal Extraction}
\ShortTitle{Fast Multi-tone Phase Calibration Signal Extraction}
\author{\textbf{Jan Wagner}\\
        Aalto University, Mets\"ahovi Radio Observatory, Kylm\"al\"a, Finland\\
        E-mail: \email{jwagner@kurp.hut.fi}}
\author{\textbf{Sergei Pogrebenko}\\
        Joint Institute for VLBI in Europe (JIVE), Dwingeloo, Netherlands\\
        E-mail: \email{pogrebenko@jive.nl} \speaker{ - speaker? tbd}}
\abstract{This paper discusses a fast algorithm for analyzing the phases of a multi-tone phase calibration signal. The multi-tone method is superior to its single-tone counterpart. The PCal signal is a wide-band frequency comb derived from an ultra-stable frequency standard which is then coupled
into the input stage of the radio astronomy receiver chain. The frequency comb allows to determine the signal delay, frequency response and phase shifts induced by the radio telescope instrumentation. The instrumentation delay is one of the several delay components that have to be compensated for in multi-telescope aperture synthesis. The frequency response also helps to monitor the instrumentation. To extract the information from the digitized baseband output of the receiver chain calls for fast numerical methods. A simple fast algorithm is presented in this paper. Two implementations are provided. One is based on precomputed look-up tables. The other is based on fast digital sinewave generators. Both lend themselves to parallel software processing on multi-core and SIMD-capable platforms as well as processing with FPGA gateware. We present optimized reference implementations for Intel/AMD, IBM Cell and nVidia GPU platforms.}
\FullConference{tbd - targeted for PoS(EXPReS09) 2009}

\begin{document}

\section{Introduction}

Phase calibration signal injection (PCal) is used to provide information necessary to align the phases of different baseband channels as a function of time, removing the effects of any differences and fluctuations in the instrumentation. Detecting and compensating the analog signal delays is important in astrometric and geodetic observations. Phase delay differences that occur at the observatories themselves, not as a part of global clock model and geometric propagation delay models, must be taken into account in VLBI correlation.

Phase calibration can detect relative phase changes over time, but there is a $2\pi$ phase wrapping ambiguity. It can be somewhat eased using absolute cable delay measurement, tracking closure phase over several baseband channels in the correlator and by using multi- instead of single-tone calibration signal injection. 
This multitone signal is injected into the topmost section of the RF chain, at the input of the LNA, separately for each polarization. This creates a common phase reference point for each baseband signal derived from the same polarization \cite{walker}.

Typical PCal pulses are separated by 1~\textmu{}s, effectively forming a frequency comb with 1~MHz spacing, see Figure 1. The pulses are derived from a local frequency reference with superior short-term stability (STS), usually a hydrogen maser or CSO oscillator. The RF chain has to be locked
to the same reference.

\begin{figure}[h]
  \centering
  \includegraphics[scale=0.33]{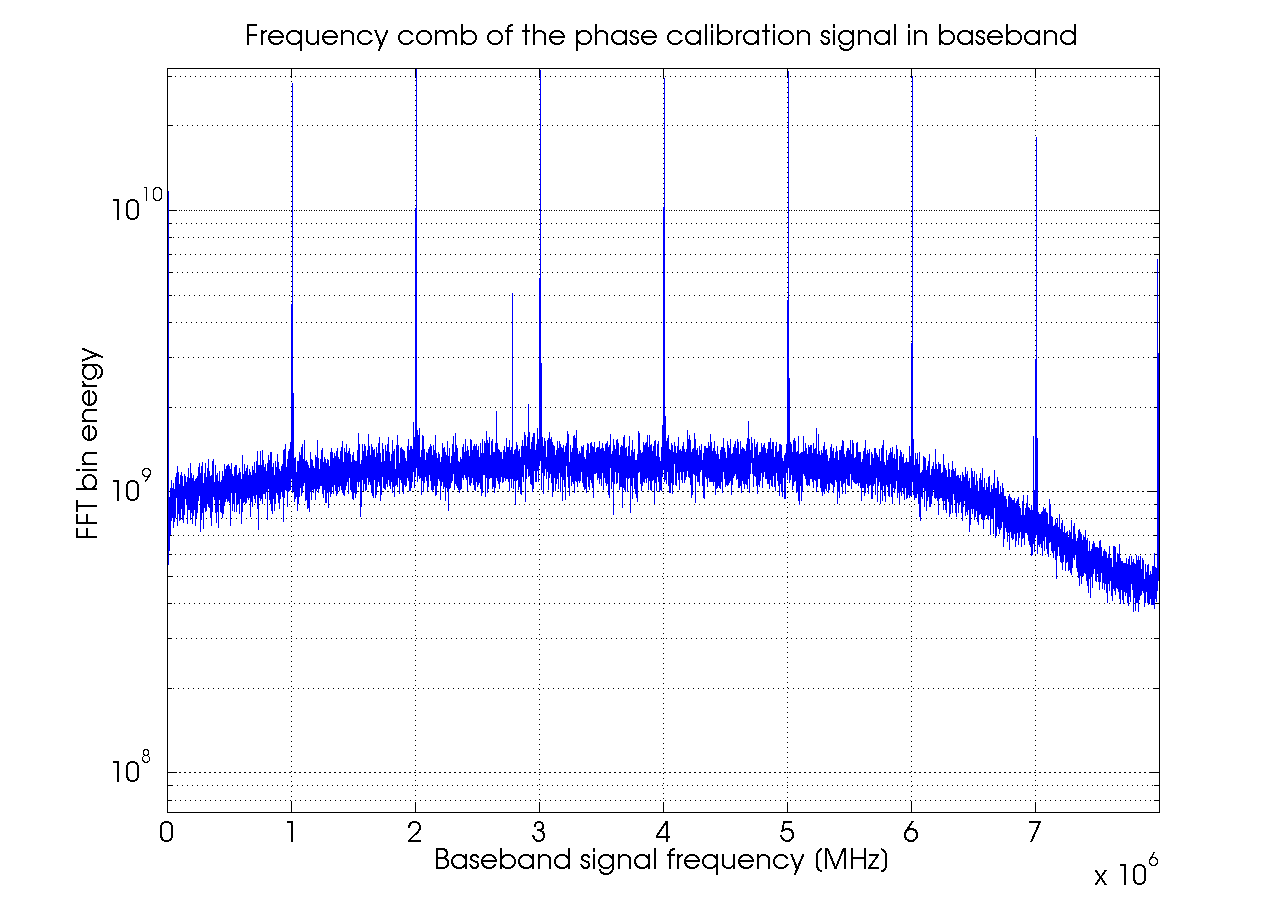}
  \caption{Frequency spectrum of a typical digitized baseband signal. The phase calibration comb has a spacing of 1 MHz and the in this case the comb offset is 0 Hz. The peak near 2.8 MHz is a spacecraft transmission.}
  \label{fig:pcalibob}
\end{figure}

VLBI data is present in digital form with all baseband channels sampled at relatively high rates (32 to 1024~MHz) and with 2-bit or higher bit resolution. Digital signal processing can be used to extract PCal tone phases from each baseband sample sequence. With an efficient implementation the PCal extraction can be performed in real-time and the online output can be used as a part of phase and delay correction during real-time correlation.

\section{Fast PCal extraction method}

The method first proposed by S. Pogrebenko works in time domain and relies on the fact that the individual
sinusoid tones in the multi-tone frequency comb have an equidistant frequency spacing $\delta f_{comb}$ \cite{pogrebenko93,pogrebenko93p2}. In radio astronomy receiver chains, this frequency comb consists of point frequencies placed at integer multiples of 1~MHz (T=1~\textmu{}s) and it covers several GHz of radio bandwidth. This comb signal is coupled into the telescope sky signal before the LNA stage of the radio receiver. With the various LO mixing frequencies that are used down the receiver chain, the comb in the output baseband signal will be shifted such that the first tone in the baseband spectrum does not reside at 0 Hz but instead at some $f_{offset}$. That is, the sky signal (usually a galaxy noise signal) in baseband (discrete $s[n]$), together with the PCal comb signal in baseband (discrete $p[n]$) can be expressed in discrete sample form as

\begin{equation}
\label{eq:pcaltdd}
x[n] = s[n] + p[n] = s[n] + \sum_{i=0}^{N-1} sin((i \cdot \hat\omega_{comb} + \hat\omega_{offset}) \cdot n + \phi_{i})
\end{equation}

where the total number of PCal point frequencies residing in the $f_{s}/2$ hertz band of the baseband signal and its mirror image is $N = f_{s}/f_{comb}$ (Figure~\ref{fig:pcalibob}: $N = 16$), the normalized $\hat\omega_{comb} = 2\pi\frac{f_{comb}}{f_{s}}$, $\hat\omega_{offset} = 2\pi\frac{f_{offset}}{f_{s}}$ and the instrumentation and delay-dependant phase shifts of each individual tone are $\phi_{i}$. For simplicity we first exclude filter-induced phase shifts and assume that phase shifts of the tones are only due to a frequency-independent time delay $t_d$, for example the propagation delay of a coaxial cable. That is, $\phi_{i}=(i \cdot \hat\omega_{comb} + \hat\omega_{offset})\cdot{t_d f_s}\forall{i}$. 

We note that if $f_{offset}=0$ and thus the comb signal has combs placed exactly at $f_{comb}$ hertz and all integer multiples thereof, when sampled at $f_{s}$, results in a sample sequence that repeats with a period of at most $N_{bins} = \frac{f_{s}}{f_{comb}}$ samples. Thus we can segment the input signal $x[n]$ into continuous non-overlapping $N_{bins}$-sample segments. As the PCal signal $p[n]$ usually has a low energy compared to the observed galaxy noise signal $s[n]$, we time-integrate several segments using direct summation:

\begin{equation}
\label{eq:pcaltdinteg}
P[i] = \sum_{j=0}^{\infty} x[i + j\cdot{N_{bins}}] \quad \text{where} \ i = 0,\dots{},N_{bins}-1 
\end{equation}

Each of the bin $i$ in $P[i], \  i = 0,\dots{},N_{bins}-1$ correspond to a time domain delay $t_d(i) = i/f_{s}$. The original average delay $t_d$ can be recovered using for example a hyperbolic estimate across the neighboring bins around the largest component. Picosecond accuracy can be reached with long integration times, or alternatively, short integration times combined with multiple passes over the $x[n]$ input data, iteratively counter-rotating the input data using the residual phase shift detected in the previous pass. 

The discrete Fourier transform $DFT(P)$ of the bins directly corresponds to the magnitude and phase response at each of the PCal comb frequencies. It is possible to deduce the combined relative instrumentation delay and a receiver filter response curve from this spectrum. We will provide data analysis examples in Section~\ref{S:danalysis} later. Figure~\ref{fig:pulseplot} in that section demonstrates a typical $P$ and deduced time delay $t_d$ as well as the $DFT(P)$ transform result. Ripple is not noise but part of the actual transfer function of station instrumentation.

Next, we note that if $f_{offset}\neq{0}$, as typically is the case in radio astronomic PCal, we can use the binning presented above if we first rotate the baseband signal spectrum by $-f_{offset}$, such that PCal tones of the rotated spectrum land at exact integer multiples of $f_{comb}$ again

\begin{equation}
\label{eq:pcaltddshifted}
\hat{x}[n] = x[n] \cdot e^{-j\hat\omega_{offset}\cdot{n}}
\end{equation}

which is then summed as in \ref{eq:pcaltdinteg}. The immediate approach for generating $\hat{x}[n]$ is to rotate every sample $x[n]$ using for example C/C++ sinf() cosf() math library functions with a running phase argument as in \ref{eq:pcaltddshifted}.

Significantly faster rotation is possible with two different approaches. For the first, we note the sampled complex rotator with $f_{offset}\neq{0}$ repeats with a shortest repeat length of

\begin{equation}
\label{eq:vrotatelen}
N_{rotator} = \frac{f_{s}}{gcd(f_{s}, f_{offset})}
\end{equation}

and this allows us to pre-compute a complex rotator vector of length $N_{rotator}$. The offsets commonly used in radio astronomy result in some hundred kilobytes and at most some megabytes of memory to store the vector. This means computation benefits from high-speed processor caches. Other choices of $f_{offset}$, however, may result in an impractically long vector and we loose the fast cache benefit.

The second approach is more flexible and only slightly more computation intensive. We note that the monochromatic rotation tone has a fixed phase increment between samples

\begin{equation}
\label{eq:vrotatearg}
\Delta\phi = 2\pi \frac{f_{offset}}{f_{s}}
\end{equation}

and instead of a precooked complex vector we can use any of the several methods available for digital sine wave generation, keeping in mind we can optimize them for a constant phase increment. That is, a fixed frequency sine oscillator is sufficient. 

Available options are Direct Digital Synthesis (DDS) and recursive approaches. Standard CORDIC is suitable when no hardware multipliers are available. With multipliers, one can use two-dimensional vector rotation, equivalent to a Coupled Form Oscillator \cite{mitra}. The modified coupled form or ``Magic Circle'' method has numerically ideal behaviour, while the computationally fastest method is the Direct Form 2nd Order Resonator (DFR). The DFR noise may be reduced using truncation error feedback if desired \cite{soanalysis}. Other derived forms of oscilaltors exist, such as the Digital Waveguide Oscillator \cite{dwo}. Recursion relations for the first three methods are

\begin{equation}
\label{eq:recursionforms}
\begin{split}
\text{Coupled Form (CFO)} & \\
y_{1}[n]& = c_n y_{1}[n-1] + s_n y_{2}[n-1] \\
y_{2}[n]& = -s_n y_{1}[n-1] + c_n y_{2}[n-1] \\
\text{Magic Circle (MCO)} & \\
y_{1}[n]& = y_{1}[n-1] + \epsilon y_{2}[n-1] \\
y_{2}[n]& = -\epsilon y_{1}[n-1] + y_{2}[n-1] \\
\text{Direct Form 2nd Order Resonator (DFR)} & \\
y_{1}[n]& = 2 c_n y_{1}[n-1] - y_{2}[n-1] \\
y_{2}[n]& = y_{1}[n-1]
\end{split}
\end{equation}

where $c_n = cos(\Delta\phi)$, $s_n = sin(\Delta\phi)$ and the magic circle parameter is $\epsilon = 2 sin(\Delta\phi/2)$. Analysis of stability, distortion and quantization effects of the above oscillators can be found in \cite{soanalysis}. For low frequencies $\hat\omega_{offset}$ the coefficient $2c_n \rightarrow 2.00^{-}$ and the quantization error distorts the DFR. The error can be reduced at cost of performance by using double precision, error feedback schemes or improved coupled forms.

With the PCal comb one may also select a higher rotator frequency $\hat\omega_{shift} = \hat\omega_{offset} + k\hat\omega_{comb}, 1 \le k < N_{bins}/2$ to reduce the error significantly. The final integrated PCal output then requires only a simple bin reordering step and data can be interpreted as before.

The DFR has a particularly fast implementation on platforms that have a Fused Multiply-Add
(FMA) instruction. A single FMA arithmetic op provides the next oscillator output value:

\begin{equation}
\label{eq:dfr}
y[n+1] = 2 cos(\Delta\phi) \cdot y[n] - y[n-1]
\end{equation}

Additionally, SIMD-capable platforms such as Intel/AMD processors with SSE instructions, IBM Cell with SPU vector units, PowerPC with AltiVec, nVidia and AMD graphics cards, should use fast vectorized Single Instruction Multiple Data (SIMD) instructions and pack four 32-bit float values into one 128-bit SIMD registers. Rewriting \ref{eq:dfr} for SIMD as follows

\begin{equation}
\label{eq:simddfr}
\begin{split}
y_{0}[n+1]& = 2 cos(2\Delta\phi) \cdot y_{0}[n] - y_{0}[n-1] \\
y_{1}[n+1]& = 2 cos(2\Delta\phi) \cdot y_{1}[n] - y_{1}[n-1] \\
y_{2}[n+1]& = 2 cos(2\Delta\phi) \cdot y_{2}[n] - y_{2}[n-1] \\
y_{3}[n+1]& = 2 cos(2\Delta\phi) \cdot y_{3}[n] - y_{3}[n-1]
\end{split}
\end{equation}

and by selecting the initial values as

\begin{equation}
\label{eq:simddfrinit}
\begin{split}
\overline{y[0]}& = [y_{0}[0], y_{1}[0], y_{2}[0], y_{3}[0]] = [cos(0), sin(0), cos(\Delta\phi), sin(\Delta\phi)] \\
\overline{y[1]}& = [cos(2\Delta\phi), sin(2\Delta\phi), cos(3\Delta\phi), sin(3\Delta\phi)]
\end{split}
\end{equation}

yields a complex-valued oscillator with two complex points ({Re,Im}) stored in one 128-bit vector. Combining \ref{eq:simddfr}, \ref{eq:simddfrinit} the single fused multiply-add instruction that computes two new complex values becomes $\overline{y[n+1]} = fma(vec4float(2 cos(2\Delta\phi)), \overline{y[n]} , \overline{y[n-1]}$).

Next we will turn to source code implementations of \ref{eq:pcaltddshifted} based on a precomputed \ref{eq:vrotatelen} $N_{rotator}$-sample vector and a fast implementation of \ref{eq:simddfr} SIMD DFR oscillator on several platforms.

\label{S:danalysis}\section{Data Analysis}

\begin{figure}[h]
  \centering
  \includegraphics[scale=0.66]{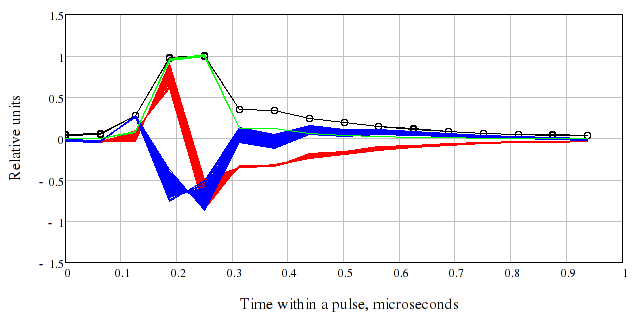}
  \caption{Overlay plot of the 112 captured pulses, 10~s integration time per pulse. Red and blue traces indicate the real and imaginary parts of the pulses, while black shows the amplitude envelope and green the power envelope.}
  \label{fig:pulseplot}
\end{figure}

Multi-tone PCal data analysis allows retrieving the essential parameters of the VLBI station instrumentation. It all can be summarized as frequency and time behaviour of the complex bandpass profile. As an example of a data set to analyse we took the data captured during the VEX spacecraft observations with Metsähovi radio telescope carried out on 2009.08.28 at X-band using standard VLBA-style data acquisition rack, MkIV formatter unit and PCEVN/VSIB data recorder. Observational bandwidth was 8~MHz and comb spacing 1~MHz. The BBC LO offset was set $f_{offset} = 10$~kHz. Continuous 19 minute scans were recorded. Captured data were processed using open-source Metsähovi SWSpec software that implements the precomputed rotator method [7]. Single integration was set to 10 seconds and 112 integrated time-integrated PCal results (pulses) were subject to further analysis which was performed on the MathCAD platform. Figure~\ref{fig:pulseplot} shows the set of the captured PCal pulses $P_i, i = 1,\dots{},112$ in overlay.

\begin{figure}[htb]
  \centering
  \includegraphics[scale=0.66]{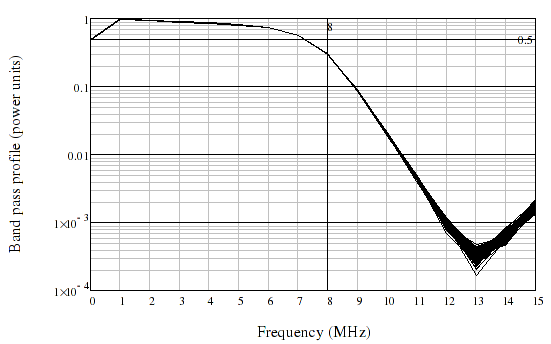}
  \caption{An overlay plot of 112 band pass profiles retrieved from the PCal pulses with 10~s integration per profile. A vertical marker indicates an 8~MHz frequency cut-off, while a horizontal marker indicates a -3~dB level.}
  \label{fig:pulseplot2}
\end{figure}

The Fourier transform $DFT(P)$ of the complex pulses gives the estimate of the instant band pass profile, as it is illustrated in Figure~\ref{fig:pulseplot2}. Note that the retrieved band pass profile covers the frequency range of 2 times the Nyquist, although the over-Nyquist and over-DC mirror bands are mixed together in the higher frequency domain.

\begin{figure}[htb]
  \centering
  \includegraphics[scale=0.66]{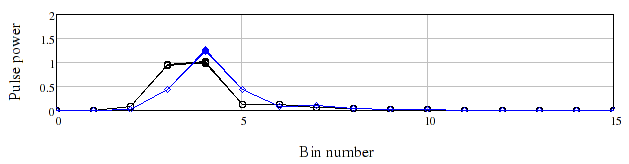}
  \caption{Original (black) and estimated group delay corrected (blue) pulses plotted in overlay.}
  \label{fig:grpdelay}
\end{figure}

The next step of the PCal data analysis is to estimate the group delay of the pulses, as the position of the maximum power. Figure~\ref{fig:grpdelay} illustrates the original pulses and group delay corrected.

\begin{figure}[htb]
  \centering
  \includegraphics[scale=0.66]{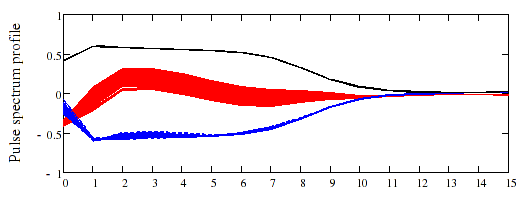}
  \includegraphics[scale=0.66]{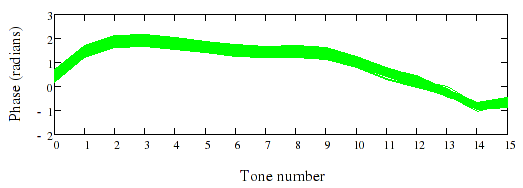}
  \caption{Group delay variation corrected band pass profiles, plotted in overlay, amplitude (black), Re/Im
(red and blue), phase (green).}
  \label{fig:correctedbpass}
\end{figure}

After the group delay is estimated, we can plot the group delay corrected band pass profile, as shown in Figure~\ref{fig:correctedbpass}.

\begin{figure}[htb]
  \centering
  \includegraphics[scale=0.66]{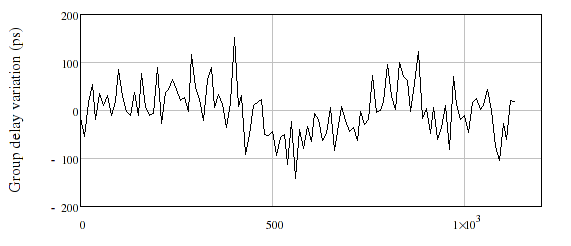}
  \includegraphics[scale=0.66]{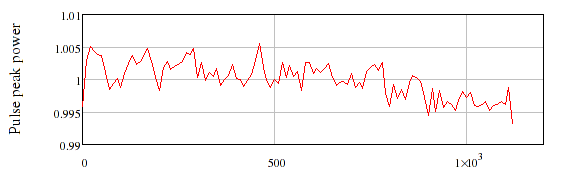}
  \includegraphics[scale=0.66]{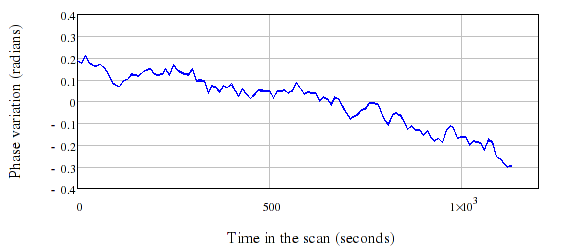}
  \caption{Time behavior of the pulse group delay (upper panel, black trace), pulse peak power (middle panel, red trace), and phases of the major tones, 1 to 7~MHz, (lower panel, blue trace). PCal integration time was 10~seconds.}
  \label{fig:pcalVsTime}
\end{figure}

Time behavior of the pulse power (related to the telescope gain), pulse group delay (related to the instrumental delay length) and phases of different spectral components are shown in Figure~\ref{fig:pcalVsTime}. The data points are taken from a time series of non-overlapping integrated PCal results. Each of these PCal result covers 10~seconds of baseband data.

\label{S:impl}\section{Reference implementations}

The reference implementation of the shifted rotation Equation \ref{eq:pcaltddshifted} with subsequent binning is provided in the listing below. This reference implementation is generally slow. In addition, performance and accuracy is prone to decline as the phase argument increases over millions of samples. The precomputed rotator and online oscillator methods do not suffer from these issues.

\begin{lstlisting}
float* samples = ...;
const size_t Nbins = f_sampling / f_comb; /* nr of tone bins */
float accu_re[Nbins] = {0.0};
float accu_im[Nbins] = {0.0};
for(size_t n=0; n<samplecount; n++) {
    accu_re[n%Nbins] += cosf(2*M_PI*(f_offset/f_s)*n) * samples[n];
    accu_im[n%Nbins] += sinf(2*M_PI*(f_offset/f_s)*n) * samples[n];
}
\end{lstlisting}

The proposed faster methods to perform Equation \ref{eq:pcaltddshifted} can be implemented in both floating-point as well as fixed-point arithmetic. For performance comparisons and for public domain reference implementations, we wrote code for precomputed and fast oscillator-based extraction methods with floating point arithmetic and vectorized for three SIMD-capable multi-core platforms, namely on the Intel/AMD x86 platform, the IBM Cell platform and the nVidia graphics card computing platform. A fixed-point implementation for FPGA parallelism is discussed further below.

\subsection{Precomputed implementation using Intel/AMD Performance Primitives}

On Intel/AMD platforms we can utilize libraries such as the Intel IPP library or the AMD Framewave library \cite{ipp,framewave}. The highest throughput with good cache locality was
achieved with the implementation below.

\begin{lstlisting}
float* samples = ...;
size_t gcd(size_t, size_t); /* extern, Greatest Common Divisor */
const size_t Nbins = f_sampling / f_comb; /* number of tone bins */
const size_t Nrotator = f_sampling / gcd(f_sampling, f_offset);

/* Precompute (can be skipped next time) */
Ipp32fc vrot[Nrotator];
for (size_t n=0; n<Nrotator; n++) {
   Ipp32f dphi = 2*M_PI*f_offset/f_sampling;
   vrot[n].re = Ipp32f(cosf(dphi * Ipp32f(n)));
   vrot[n].im = Ipp32f(sinf(dphi * Ipp32f(n)));
}

/* Intel IPP quick-ref:                           */
/*   ippsMul_type(src,src,dst,len) : out-of-place */
/*   ippsAdd_type_I(src,src_dst,len) : in-place  */

/* Clear the bins */
Ipp32fc accu[Nbins];
ippsZero_32fc(accu, Nbins);

/* Process all data: segment it into Nrotator-sized pieces,   */
/* then rotate, finally fold Nrotator-sized vector into Nbins */
Ipp32f* segment = (Ipp32f*)samples;
Ipp32fc tmp[Nrotator];
for (size_t s=0; s<(samplecount/Nrotator); s++) {
  ippsMul_32f32fc(segment, vrot, tmp, Nrotator); 
  Ipp32fc* pulse = tmp;
  for (size_t pp=0; pp<(Nrotator/Nbins); pp++) {
      ippsAdd_32fc_I(pulse, accu, Nbins); 
      pulse += Nbins;
  }
  segment += Nrotator;
}
\end{lstlisting}

When the $N_{rotator}$ vector is short, with sufficient top-level cache available, throughput could be improved further by accumulating into a $N_{rotator}$-sample accumulator and segmenting and folding the long accumulator into the final PCal $N_{bins}$ bins later, outside the samplecount loop. At the moment however a ippsAddProduct\_32f32fc\_I() is not part of Intel IPP; it can be programmed
with SSE2 instructions though.

\subsection{Oscillator implementation using Intel/AMD SSE2 instructions}

The oscillator version could be implemented using the same Intel IPP library intrinsics, but the performance is not good since the library functions like to operate on long set of data vectors instead of a single vector (the complex oscillator state). To achieve highest performance, \ref{eq:simddfr}, \ref{eq:simddfrinit} should be written using Intel Streaming SIMD Extensions, in particular, SSE2 assembly instructions or their corresponding C/C++ instrinsics. The Microsoft MSDN offers an excellent overview of MMX and SSE intrinsics \cite{mssse}. 

The SSE instructions do not include a Fused-Multiply-Add, but Intel and AMD roadmaps plan to add FMA in 2011 as a part of the AVX advanced vector instruction set.

\begin{lstlisting}
#include <xmmintrin.h>
const size_t Nbins = f_sampling / f_comb; /* number of tone bins */

float* samples = ...;
float dphi = 2*M_PI*f_offset/f_sampling;
float accu[2*Nbins] __attribute__((aligned(16))) = {0.0};

__m128 mmxOldComplex_0, mmxOldComplex_1, mmxMult;
mmxOldComplex_0 = _mm_set_ps( sinf(1*dphi), cosf(1*dphi), 
                              sinf(0*dphi), cosf(0*dphi) );
mmxOldComplex_1 = _mm_set_ps( sinf(3*dphi), cosf(3*dphi), 
                              sinf(2*dphi), cosf(2*dphi) );
mmxMult = _mm_set1_ps( 2*cosf(2*dphi) );

assert((samplecount%(4*Nbins)) == 0);
for (size_t s=0; s<samplecount; s += Nbins ) {
   __m128 mmxData, mmxPcal, mmxTmp, mmxLoad;
   for (int bin=0; bin<Nbins; bin+=4, samples+=4) {
       /* rotate and bin first 2 samples */
       mmxLoad = _mm_load_ps(data);  // or _mm_loadu_ps()
       mmxData = _mm_unpacklo_ps(mmxLoad, mmxLoad);
       mmxPcal = _mm_load_ps(&accu[2*bin+0]);
       mmxData = _mm_mul_ps(mmxData, mmxOldComplex_0);
       mmxPcal = _mm_add_ps(mmxPcal, mmxData);
       _mm_store_ps(&accu[2*bin+0], mmxPcal);
       /* rotate and bin next 2 samples */
       mmxData = _mm_unpackhi_ps(mmxLoad, mmxLoad);
       mmxPcal = _mm_load_ps(&accu[2*bin+4]);
       mmxData = _mm_mul_ps(mmxData, mmxOldComplex_1);
       mmxPcal = _mm_add_ps(mmxPcal, mmxData);
       _mm_store_ps(&accu[2*bin+4], mmxPcal);
       /* compute next four complex value pairs */
       mmxTmp = _mm_mul_ps(mmxMult, mmxOldComplex_1);
       mmxTmp = _mm_sub_ps(mmxTmp, mmxOldComplex_0);
       mmxOldComplex_0 = mmxOldComplex_1;
       mmxOldComplex_1 = mmxTmp;
       mmxTmp = _mm_mul_ps(mmxMult, mmxOldComplex_1);
       mmxTmp = _mm_sub_ps(mmxTmp, mmxOldComplex_0);
       mmxOldComplex_0 = mmxOldComplex_1;
       mmxOldComplex_1 = mmxTmp;
   }
}
\end{lstlisting}

\subsection{Precomputed implementation using IBM Cell SPU instrinsics}

The Cell SPU cores differ from Intel Nehalem SSE units mainly in that the SPU offers bit extended data shuffling instructions and a fused multiply accumulate instruction. The instructions are documented in detail in the IBM Cell Hardware reference. A comprehensive overview of C intrinsics can be found in the Cell C/C++ Language Extensions manual\cite{spuintr}.

The Cell floating point units have certain some limitations. They do not fully support IEEE-754. In particular, denormalized numbers are not supported and rounding uses simple ``round to zero''. The latter causes some loss of precision that one has to live with \cite{cellaccu}.

The following constants are required for the spu\_shuffle instruction. They “shuffle” a vector of 4 floats [a b c d] into [a a b b] and [c c d d] respectively.

\begin{lstlisting}
const vector unsigned char _duplicate_lower = (const vector unsigned char )
   { 0x00, 0x01, 0x02, 0x03, 0x00, 0x01, 0x02, 0x03,
     0x04, 0x05, 0x06, 0x07, 0x04, 0x05, 0x06, 0x07 };
const vector unsigned char _duplicate_upper = (const vector unsigned char )
   { 0x08, 0x09, 0x0A, 0x0B, 0x08, 0x09, 0x0A, 0x0B,
     0x0C, 0x0D, 0x0E, 0x0F, 0x0C, 0x0D, 0x0E, 0x0F };
\end{lstlisting}

The precomputed implementation for Cell SPUs is below. It processes 4 input samples i.e. one input vector at a time. Use the \emph{-funroll-loops} compiler flag.

\begin{lstlisting}
/* precomputed init */
int i;
for (i =0; i <PRECOMPLEN; i++) {
   precomputed[2*i+0] = cosf(i*_dphi);
   precomputed[2*i+1] = sinf(i*_dphi);
}
/* PCal extraction with lookup */
int i, p, b;
vector float vecsmp = (vector float*)&samples[0];
vector float vecpcal= (vector float*)&pcal[0];
vector float vecprecomp = (vector float*)&precomputed[0] ;
for (i=0; i<(CHUNKSZ/4); ) {
   for (p=0; p<(PRECOMPLEN/4); ) {
      for (b=0; b<(NUMBINS/2); b+=2, p+=2, i++) {
         vector float s = vecsmp[i];
         vector float s12 = spu_shuffle(s, s, _duplicate_lower);
         vector float s34 = spu_shuffle(s, s, _duplicate_upper);
         vecpcal[b]   = spu_madd (vecprecomp[p],   s12, vecpcal[b]);
         vecpcal[b+1] = spu_madd (vecprecomp[p+1], s34, vecpcal[b+1]);
      }
   }
}
\end{lstlisting}

\subsection{Oscillator implementation using IBM Cell SPU instrinsics}

The oscillator implementation is provided below. Use the \emph{-funroll-loops} compiler flag.

\begin{lstlisting}
/* oscillator init */
float dphi = 2*M_PI * (f_offset + 0*f_comb) / fs;
vector float sc_1 = (vector float){ cosf(2*_dphi), sinf(2*_dphi),
                                    cosf(3*_dphi), sinf(3*_dphi) };
vector float sc_0 = (vector float){ cosf(0*_dphi), sinf(0*_dphi),
                                    cosf(1*_dphi), sinf(1*_dphi) };
vector float step = spu_splats(2*cosf(2*_dphi));

/* PCal extraction with DFR */
int i, b;
vector float* vecsmp =  (vector float*)&samples[0];
vector float* vecpcal = (vector float*)&pcal[0];
for (i=0; i<(CHUNKSZ/4); ) {
   for (b=0; b<(NUMBINS/2); b+=2) {
       /* get 4 samples, rotate, bin into 4 complex points */
       vector float s = vecsmp[i++];
       vector float s12 = spu_shuffle(s, s, _duplicate_lower);
       vector float s34 = spu_shuffle(s, s, _duplicate_upper);
       vecpcal[b]   = spu_madd(sc_0, s12, vecpcal[b]);
       vecpcal[b+1] = spu_madd(sc_1, s34, vecpcal[b+1]);
       / * compute next two oscillator values * /
       vector float tmp = sc_1;
       sc_1 = spu_msub(step, sc_1, sc_0);
       sc_0 = tmp;
       tmp = sc_1;
       sc_1 = spu_msub(step, sc_1, sc_0);
       sc_0 = tmp;
   }
}
\end{lstlisting}

\subsection{Precomputed implementation using nVidia CUDA}

With the CUDA SDK by nVidia gaming graphics cards by the same company can be used
for computation. The basic non-precomputed implementation is provided below. On a GTX~285, using
8388608 samples and memory mapped access, the throughput of the basic code reaches only 160
Ms/s.

\begin{lstlisting}
__global__ void pcalGPU(float *a, float *b, float *c, float start, int N)
{
   int idx = blockIdx.x * blockDim.x + threadIdx.x;
   int bin = (idx % 1600); /* 1600 = f_s / gcd(f_s, f_offset) */
   float w = 0.00392699081698724; /*  w= 2*PI*f_offset/f_s */
   float phi = start + (w*idx);
   if (idx<N) {
      a[bin] += __cosf(phi) * c[idx];
      b[bin] += __sinf(phi) * c[idx];
   }
   /* host CPU shouldcollapse 1600 points into 16 bins
      at end of integration */
}

dim3 block(256);
dim3 grid((unsigned int)ceil(samplecount/(float)block.x));
pcalGPU<<<grid, block>>>(d_a, d_b, d_c, 0.0f, samplecount);
\end{lstlisting}

A four times faster version is possible with just three changes. In the implementation below,
Stream Processors accumulate locally first. Samples are loaded from a read-only texture. Finally,
each thread processes more than one sample.

\begin{lstlisting}
texture <float, 1, cudaReadModeElementType> samplevecTex;
#define N_THREADS_PER_BLOCK 64
#define N_SAMPS_PER_THREAD 2048
#define N_BINS 8

__global__ void pcalGPU(float * out, unsigned int N)
{
   __shared__ float pulsesRe[N_THREADS_PER_BLOCK][N_BINS];
   __shared__ float pulsesIm[N_THREADS_PER_BLOCK][N_BINS];
   int bin=0, rvecIdx=0, samplenr=0;
   unsigned int sample_idx = N_SAMPS_PER_THREAD *
      (threadIdx.x + blockIdx.x * blockDim.x);

   / * Accumulate to local registers / memory */
   while ((samplenr < N_PRECOMPUTED) && (sample_idx < N)) {
       float rotRe = __sinf(sample_idx * 2*M_PI*10e3/16e6);
       float rotIm = __cosf(sample_idx * 2*M_PI*10e3/16e6);
       bin = (int)(sample_idx % N_BINS);
       float s = tex1Dfetch(samplevecTex, sample_idx);
       pulsesRe[threadIdx.x][bin] += s * rotRe;
       pulsesIm[threadIdx.x][bin] += s * rotIm;
       sample_idx++; samplenr++;
   }

   / * Write results to card RAM * /
   rvecIdx = 2 * N_BINS * (threadIdx.x + blockIdx.x * blockDim.x);
   for (bin=0; bin<N_BINS; bin++) {
      out[rvecIdx++] = pulsesRe[threadIdx.x][bin];
      out[rvecIdx++] = pulsesIm[threadIdx.x][bin];
   }
}

dim3 block (N_THREADS_PER_BLOCK, 1, 1);
dim3 grid(ceil(samplecount/(N_SAMPS_PER_THREAD*(float)block.x)), 1, 1);
pcalGPU<<<grid, block>>>(d_results, samplecount);
/* results need final folding on CPU * /
\end{lstlisting}

\subsection{Precomputed implementation using nVidia CUDA}

The precomputed method does not significantly differ from the direct method implementation. The complex rotator values are simply loaded from a new texture.

Like with the improved direct method, for fastest performance the number of blocks should match that of available stream multiprocessors. Threads per block should be selected to be close to the maximum number supported by the GPU. On a GTX~285 this means a total of $30\cdot{512}=15360$ parallel threads. Experimentation gives the optimal value. Each thread runs the precomputed rotator code and accumulates into local memory. At the end, the thread accumulated results are written back into mapped main memory.

At the end of a full data pass, the CPU should fold the sub-results into a final $N_{bins}$-bin PCal vector. It is possible to perform this step on the GPU, too, but the code for thread synchronizations or binary tree stepping is too long to reproduce here.

\begin{lstlisting}
texture<float,1,cudaReadModeElementType> rvecTex;
texture<float,1,cudaReadModeElementType> samplevecTex;
#define N_THREADS_PER_BLOCK 64
#define N_PRECOMPUTED 1600
#define N_BINS 8

__global__ void pcalGPU(float* out, unsigned int N)
{
   __shared__ float rotRe, rotIm;
   __shared__ float pulsesRe[N_THREADS_PER_BLOCK][N_BINS];
   __shared__ float pulsesIm[N_THREADS_PER_BLOCK][N_BINS];
   unsigned int sample_idx = N_PRECOMPUTED *
      (threadIdx.x + blockIdx.x*blockDim.x*N_PRECOMPUTED);
   int rvecIdx = 0; // = sample_idx % N_PRECOMPUTED=0
   int bin = 0; // = sample_idx % N_BINS=0
   int samplenr = 0;

   /*Accumulate to local registers/memory */
   while ((samplenr<N_PRECOMPUTED) && (sample_idx<N)) {
      float x = tex1Dfetch(samplevecTex, sample_idx);
      sample_idx++;samplenr++;
      if (threadIdx.x==0) {
         rotRe = tex1Dfetch(rvecTex, 2*rvecIdx+0);
         rotIm = tex1Dfetch(rvecTex, 2*rvecIdx+1);
       }
       pulsesRe[threadIdx.x][bin] += x*rotRe;
       pulsesIm[threadIdx.x][bin] += x*rotIm;
       rvecIdx = (rvecIdx+1) % N_PRECOMPUTED;
       bin = (bin+1) % N_BINS;
   }

   /* Write results to card RAM */
   rvecIdx = 2*N_BINS*(threadIdx.x + blockIdx.x*blockDim.x);
   for (bin=0; bin<N_BINS; bin++) {
      out[rvecIdx++] = pulsesRe[threadIdx.x][bin];
      out[rvecIdx++] = pulsesIm[threadIdx.x][bin];
   }
}
dim3 block(N_THREADS_PER_BLOCK,1,1);
dim3 grid(ceil(samplecount/(N_PRECOMPUTED*(float)block.x)), 1, 1);
pcalGPU<<<grid,block>>>(d_results, samplecount);
/* results need final folding on CPU */
\end{lstlisting}

\subsection{Oscillator implementation using nVidia CUDA}

The newest GPUs support double-precision. A numerically more accurate oscillator can be
implemented in double-precision arithmetic. The single-precision example is provided below. Like
in the precomputed method on the GPU, the sub-results delivered by the oscillator code have to be
folded by the CPU later into a final $N_{bins}$-bin PCal vector.

\begin{lstlisting}
texture<float,1,cudaReadModeElementType> samplevecTex;
#define N_THREADS_PER_BLOCK 64
#define N_SAMPS_PER_THREAD 1600
#define N_BINS 8

__global__ void pcalGPU(float* samples, float* out, unsigned int N)
{
   float rotRe_1, rotIm_1, tco, rotRe_0, rotIm_0, phi;
   __shared__ float pulsesRe[N_THREADS_PER_BLOCK][N_BINS];
   __shared__ float pulsesIm[N_THREADS_PER_BLOCK][N_BINS];
   unsigned int sample_idx = N_SAMPS_PER_THREAD*
      (threadIdx.x+blockIdx.x*blockDim.x);
   int rvecIdx = sample_idx%N_SAMPS_PER_THREAD;
   int bin = sample_idx%N_BINS;
   int samplenr = 0;
   phi = 2*M_PI*10e3/16e6;
   tco = 2*__cosf(phi);
   rotRe_1 = __cosf(phi*(sample_idx+1));
   rotRe_0 = __cosf(phi*(sample_idx));
   rotIm_1 = __sinf(phi*(sample_idx+1));
   rotIm_0 = __sinf(phi*(sample_idx));
   
   /*Accumulate to local registers/memory */
   while((samplenr<N_SAMPS_PER_THREAD)&&(sample_idx<N)){
      float tmp, x = tex1Dfetch(samplevecTex,sample_idx);
      sample_idx++; samplenr++;
      pulsesRe[threadIdx.x][bin] += x*rotRe_0;
      pulsesIm[threadIdx.x][bin] += x*rotIm_0;
      bin = (bin+1)%N_BINS;
      tmp = rotRe_1;
      rotRe_1 = tco*rotRe_1 - rotRe_0;
      rotRe_0 = tmp;
      tmp = rotIm_1;
      rotIm_1 = tco*rotIm_1 - rotIm_0;
      rotIm_0 = tmp;
   }

   /*WriteresultstocardRAM*/
   rvecIdx = 2*N_BINS*(threadIdx.x+blockIdx.x*blockDim.x);
   for(bin=0; bin<N_BINS; bin++){
      out[rvecIdx++] = pulsesRe[threadIdx.x][bin];
      out[rvecIdx++] = pulsesIm[threadIdx.x][bin];
   }
}

dim3 block(N_THREADS_PER_BLOCK, 1, 1);
dim3 grid(ceil(samplecount/(N_SAMPS_PER_THREAD*(float )block.x)), 1, 1);
pcalGPU<<<grid,block>>>(d_results, samplecount);
/* results need final folding on CPU */
\end{lstlisting}

\subsection{FPGA implementation}

Most medium- to high-end FPGAs today incorporate a set of hardware multipliers. The Xilinx Virtex-5 LX330 for example has 192~DSP48E slices each with one 18x25 -bit hardware multiplier and accumulator. Floating point IP cores may be synthesized on FPGA  but consume a high amount of resources. On a Virtex-4 a single Xilinx FPU IP core consumes around 30\% of FPGA resources.

To conserve resources for other data processinga fixed-point implementation can be preferred. Even a 18-bit fixed point oscillator implementation may still provide low enough noise and THD to be useful for PCal extraction \cite{soanalysis}. Two  DSP48(E) slices are sufficient to implement a pipelined version of \ref{eq:dfr} that computes the real and complex parts of the rotator. Two additional DSP48(E) slices are required to rotate the input sample. Bins and accumulator registers are directly implemented with general slices e.g. sufficiently long counters. 

DSP48E can be clocked in the vicinity of 550~MHz. Without further parallelism a four-DSP48E implementation can thus process~550 Ms/s. With suitable segmentation of the input sample stream and use of further DSP48E slices the throughput can be increased to several 10~Gs/s.

An actual VHDL design and its performance on a FPGA are not discussed here.

\section{Results}

The numerically most accurate method is the precomputed rotator method. Figure~7 shows
the errors of direct calculation and oscillator methods as compared to the precomputed rotator over
several million cycles.

Performance tests were run on available platforms (Intel, Cell, GPU) using $N_{bins} = 16$, $N_{rotator} = 1600$, $N_{samples} = 16 \cdot 10^6$ and a sufficient number of iterations to run longer than five seconds. We tested three Intel platform computers: a Core~i7~920 2.67~GHz in Boost Mode running kernel 2.6.28 x86\_64, a Core~2 Quad Q9550 2.83~GHz under 2.6.27 i686 and a Core~2 Quad Q8300 2.5~GHz under 2.6.28 i686. For the Cell platform we used a PlayStation~3 running at 3.1~GHz under 2.6.24. For GPU tests we used an XFX GTX285 XXX card. Table~5 shows the performance of the three extraction methods on these platforms.

\begin{table}[tbh]
\label{tbl:results}
\caption{Throughput measured as floating point input samples $x[n]$ processed per second. For the CPU processors the values are for a single utilized CPU core. For GPU values are for the entire graphics processor with memory copy overheads included.}
\begin{tabular}{| l | c | c | c |}
\hline
                          & Direct     & Precomputed & Oscillator \\ \hline
Core i7 (one core)        & 2.5 Ms/s   & 600 Ms/s    & 774 Ms/s   \\ \hline
Core2 Q9550 (one core)    & 15.5 Ms/s  & 633 Ms/s    & 656 Ms/s   \\ \hline
Core2 Q8300 (one core)    & 13.2 Ms/s  & 302 Ms/s    & 570 Ms/s   \\ \hline
IBM Cell (one SPU)        & 14 Ms/s    & 405 Ms/s    & 810 Ms/s   \\ \hline
nVidia GTX285 (240 cores) & 140 Ms/s (550 Ms/s) & 570 Ms/s & 570 Ms/s \\ 
\hline
\end{tabular}
\end{table}

On Core i7, single-precision 64-bit C library trigonometric functions were extremely slow (2.5~Ms/s). The throughput increased to 15~Ms/s with double-precision functions.

The direct method on Cell used C library trigonometric functions for 14~Ms/s and libgmath library sin14\_f() for 58~M/s. The libgmath functions calculate four values in parallel. These are only 14-bit -accurate. The speed of the closest Intel IPP 11-bit counterpart ippsSinCos\_32f\_A11 and the 24-bit -accurate ippsSinCos\_64f\_A24 were not tested on Core i7 nor Core2.

On the GPU, there are a number of bottlenecks. The main bottleneck is the CPU-side result memory that is mapped into the GPU address space to skip further memcpy instructions. Rates over 800~Ms/s were possible when results remained on GPU memory instead. Analysis of GPU memory access patterns showed the patterns are not optimal. Improving this side of the code may result in throughput in the Gs/s range.

\section{Conclusions}

Multi-tone phase calibration signal extraction significantly improves the quality of instrumental delay/phase calibration with respect to single tone techniques. We presented a fast multi-tone extraction algorithm with two possible high-throughput software implementations. Both lend themselves to SIMD vectorization and multi-core processing. Example implementations for Intel/AMD, IBM Cell and nVidia CUDA computing platforms were presented. Performance benchmarks showed that typical performance exceeded 500~Ms/s on a single core. A potential FPGA implementation was discussed.

Of the two software implementations, the first uses a precomputed complex sinusoid vector and requires very few arithmetic operations, but also demands access to fast memory. While the method is very fast for those comb offset frequencies that are typically used in VLBI radio astronomy at the moment – offsets are commonly chosen to be some multiple of 10 kHz – the storage requirements of the precomputed vector grow if the chosen comb offset less optimal. In some situations the precomputed vector may exceed the size of available fast cache memory. Prime valued frequency offsets are a pathological example. When the precomputed vector no longer fits into the cache, an unlikely scenario in current VLBI, the performance becomes bandwidth-limited and may take a severe hit.

The second software implementation uses processor registers to compute the complex sinusoid using a fast second-order digital oscillator. This method does not require additional memory and uses only few more arithmetic operations compared to the first method. Performance is arithmetic-limited. On every platform presented in this paper the second method was faster than the precomputed method. It comes at the cost of lower accuracy over longer integration times. The method sets no restrictions on the possible comb offset frequency.

In conclusion, the first presented method is suitable for platforms where memory is faster than floating-point arithmetic. The second method is suitable on platforms where floating-point arithmetic is particularly fast, also provided that the accuracy trade-off is acceptable. The second method has flexibility advantages over the first method. Both methods are provided to the public domain. The presented source code is provided under GNU GPL v2 licensing.

\newpage

\end{document}